 \documentclass[smallabstract,smallcaptions]{dccpaper}

\usepackage{epsfig}
\usepackage{amsmath}
\usepackage{amssymb}
\usepackage{color}
\usepackage{url}
\usepackage{booktabs}

\newlength{\figurewidth}
\newlength{\smallfigurewidth}

\begin{document}

\title
{\large
\textbf{Audio-Visual Cross-Modal Compression for Generative Face Video Coding}
}

\author{%
Youmin Xu$^{\ddag\S}$, Mengxi Guo$^{\dag\S}$\thanks{$^{\S}$Equal contribution.}, Shijie Zhao$^{\dag\ast}$\thanks{$^{\ast}$Corresponding author.}, Weiqi Li$^{\ddag}$, Junlin Li$^{\dag}$, Li Zhang$^{\dag}$, Jian Zhang$^{\ddag}$\\[1em]
{\small\begin{minipage}{\linewidth}\begin{center}
\begin{tabular}{c}
$^{\dag}$Bytedance Inc., Shenzhen, China \& San Diego, CA, USA \\
$^{\ddag}$School of Electronic and Computer Engineering, Peking University \\[0.5em]
\texttt{\{guomengxi.qolab, zhaoshijie.0526, lijunlin.li, lizhang.idm\}@bytedance.com} \\
\texttt{youmin.xu@stu.pku.edu.cn, liweiqi@stu.pku.edu.cn, zhangjian.sz@pku.edu.cn}
\end{tabular}
\end{center}\end{minipage}}
}



\maketitle
\thispagestyle{empty}

\begin{abstract}
Generative face video coding (GFVC) is vital for modern applications like video conferencing, yet existing methods primarily focus on video motion while neglecting the significant bitrate contribution of audio. Despite the well-established correlation between audio and lip movements, this cross-modal coherence has not been systematically exploited for compression. To address this, we propose an \textbf{Audio-Visual Cross-Modal Compression (AVCC)} framework that jointly compresses audio and video streams. Our framework extracts motion information from video and tokenizes audio features, then aligns them through a unified audio-video diffusion process. This allows synchronized reconstruction of both modalities from a shared representation. In extremely low-rate scenarios, AVCC can even reconstruct one modality from the other. Experiments show that AVCC significantly outperforms the Versatile Video Coding (VVC) standard and state-of-the-art GFVC schemes in rate-distortion performance, paving the way for more efficient multimodal communication systems.
\end{abstract}

\begin{figure*}[t]
\includegraphics[width=0.91\textwidth]{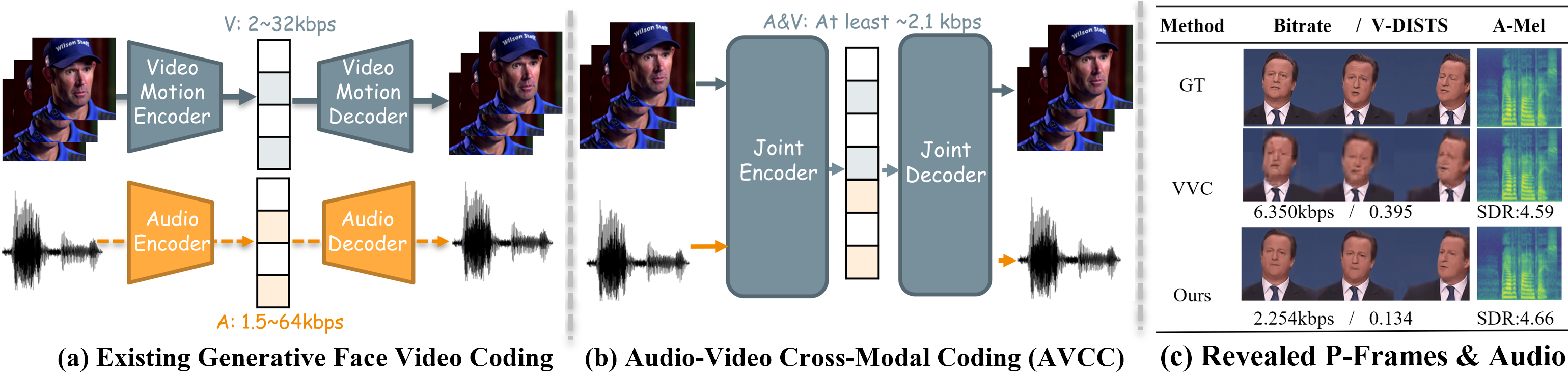}
\vspace{-0.3cm}
\caption{Comparison between (a) existing generative face video coding, which processes audio and video independently, and (b) our proposed AVCC framework. AVCC constructs a mutual representation from audio and video temporal features to assist the joint encoding and decoding of both streams.}
\label{fig:teaser}
\vspace{0.5em}
\end{figure*}

\begin{figure*}[t]
\centering
\centerline{\includegraphics[width=0.92\linewidth]{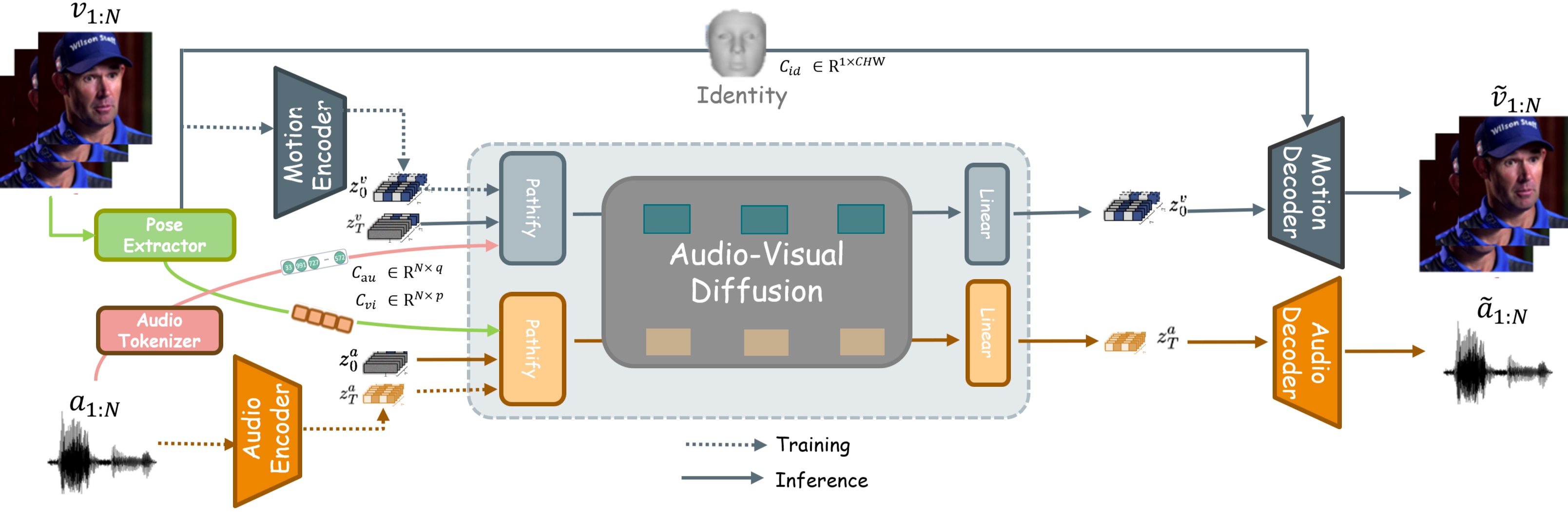}}
\caption{Overall framework of our proposed AVCC method. The encoder jointly compresses audio and video inputs using cross-modal information. The decoder uses a unified diffusion process to reconstruct synchronized audio and motion vectors, which drive the generation of video frames from a transmitted key frame.}
\label{fig:totalframe}
\vspace{-1em}
\end{figure*}

\Section{Introduction}

The proliferation of internet video has made efficient compression critical. While traditional codecs like H.264/AVC~\cite{sullivan2012overview} and VVC~\cite{bross2021overview} operate at the signal level, recent generative video coding (GVC) methods leverage semantic representations to achieve superior rate-distortion (RD) performance, especially for face videos which exhibit strong prior patterns~\cite{chen2024generative}. Current Generative Face Video Coding (GFVC) frameworks~\cite{FOMM, wang2021Nvidia, chen2023csvt} transform videos into compact representations like keypoints or features, which are then used by a decoder to reconstruct realistic videos.

However, these methods overlook a critical aspect of ultra-low bitrate communication: the audio stream. Advanced audio codecs~\cite{van2017neural, zeghidour2021soundstream} have reduced audio bitrates, but they still constitute a significant portion of the total bandwidth. Existing GFVC frameworks compress audio and video independently, failing to exploit the strong inherent correlation between them. Research in lip-reading~\cite{assael2016lipnet} and talking-head generation~\cite{chen2023dae} has demonstrated that lip movements can infer speech and vice versa.

To bridge this gap, we introduce \textbf{Audio-Visual Cross-modal Compression (AVCC)}, a novel framework for joint audio-visual compression. As shown in Fig.~\ref{fig:teaser}, AVCC extracts motion information from video and tokenized features from audio. These representations are then fused within a unified \textbf{Audio-Visual Diffusion (AVDiff)} model, which generates synchronized audio and video outputs from noise. By exploiting this cross-modal redundancy, AVCC achieves superior compression efficiency compared to independent coding methods. Our main contributions are:
\begin{itemize}
    \item We propose AVCC, the first GFVC framework to integrate audio and video into a unified compression process, leveraging their temporal correlation to enhance efficiency.
    \item We introduce the AVDiff model, which effectively aligns and reconstructs audio and video streams from a shared latent space, enabling high-quality synthesis even at very low bitrates.
    \item Our experiments show that AVCC significantly outperforms VVC and other state-of-the-art GFVC methods, demonstrating the value of cross-modal compression for next-generation communication systems.
\end{itemize}

\Section{Related Works}
\SubSection{Face Video Coding}
Generative video coding diverges from conventional hybrid coding frameworks by focusing on efficient, flexible data representation. Drawing inspiration from early Model-Based Coding~\cite{7268565}, these approaches emphasize the extraction of compact spatial and temporal representations. Unlike early methods, current GFVC techniques are largely built upon advanced face animation and reenactment models~\cite{FOMM,hong2022depth}, relying heavily on the powerful inference capabilities of deep learning to ensure high-quality video reconstruction~\cite{rombach2022high}. While compact representations are efficient, they may offer fixed-dimensional outputs that lack adaptability to fluctuating network conditions, reducing coding flexibility~\cite{ming2024survey}.

\SubSection{Audio-Visual Representation}
Cross-modal data compression focuses on removing redundancy between different media types. The common approach is to project data into a shared semantic space for a unified, compact representation. Audio-Visual Hidden Unit BERT (AV-HuBERT)~\cite{shi2022avhubert} is a self-supervised model designed to learn robust representations from both audio and visual data. It processes visual input from the mouth region and audio input as MFCCs, fusing them using transformer blocks to learn comprehensive audio-visual speech representations. Compared to earlier models like SyncNet~\cite{prajwal2020lip}, AV-HuBERT provides more consistent performance, making it a reliable choice for high-quality audio-visual synchronization, which inspires our joint compression approach.

\Section{Methods}

\SubSection{Overall Encoding/Decoding Processes}

As illustrated in Fig.~\ref{fig:totalframe}, our AVCC framework encodes video and audio into a unified bitstream containing three components: identity codes ($C_{id}$), pose codes ($C_{vi}$), and audio tokens ($C_{au}$).

\noindent\textbf{Encoding:} The identity code $C_{id}$ is extracted from the first frame (I-frame) to represent static facial features. For subsequent frames (P-frames), a \textbf{Pose-Extractor} computes motion parameters, which are encoded as the pose code $C_{vi}$. Concurrently, an \textbf{Audio Tokenizer} converts the audio stream into discrete audio tokens $C_{au}$.

\noindent\textbf{Decoding:} The decoder uses the identity code $C_{id}$ to initialize the facial structure. The pose code $C_{vi}$ and audio tokens $C_{au}$ are fed into our \textbf{Audio-Visual Diffusion (AVDiff)} model. This model jointly reconstructs the audio signal and the dense motion field, which is then used to animate the initial identity frame and generate the final video output. This joint process ensures synchronization and leverages cross-modal information for higher quality reconstruction.

\SubSection{Audio Tokenizer and Video Pose-Extractor}
\label{sec:v2a_and_ldmdec}

\noindent\textbf{Audio Tokenizer.} Our audio encoding approach employs a dual-encoder architecture to capture diverse audio features efficiently. The semantic encoder, based on a self-supervised Audio Masked Autoencoder (AudioMAE), captures high-level audio features. These features are then quantized into discrete semantic tokens using an ensemble k-means clustering approach. An acoustic encoder complements this by capturing residual audio details not represented by the semantic features. The outputs of both encoders are concatenated, forming a comprehensive representation that is converted into discrete indices. For reconstruction, a diffusion-model-based decoder is employed to generate clean audio conditioned on these indices.

\noindent\textbf{Video Pose-Extractor.} Our Pose-Extractor encodes static identity and dynamic motion separately. From the I-frame, it extracts the identity code $C_{id}$. For P-frames, it computes motion parameters describing head pose, lip movement, and eye motion, encoding them into the pose code $C_{vi}$. We employ a latent image animator (LIA) that decomposes motion into identity-specific and motion-specific latent codes, allowing for a more robust and compact representation of facial dynamics.

\SubSection{Audio-Visual Diffusion for Joint Decoding}
\label{avdiff}

The core of our decoder is the AVDiff model, which achieves temporal multimodal alignment by jointly generating audio and video from their compressed representations. We adapt a Diffusion Transformer (DiT) architecture for this task. Starting from Gaussian noise, the model iteratively denoises the latent representations of audio ($z_a$) and video ($z_v$). Specifically, given paired noise $(z_{a}^T, z_{v}^T)$, a joint denoising network $\theta_{av}$ models the reverse process:
\begin{align}
p_{\theta_{av}}(z_{a}^{t - 1}|z_{a}^{t}, z_{v}^{t}) &= \mathcal{N}(z_{a}^{t - 1}; \mu_{\theta_{av}}(z_{a}^{t}, z_{v}^{t}, t), \sigma_{t}^{2}I), \\
p_{\theta_{av}}(z_{v}^{t - 1}|z_{v}^{t}, z_{a}^{t}) &= \mathcal{N}(z_{v}^{t - 1}; \mu_{\theta_{av}}(z_{v}^{t}, z_{a}^{t}, t), \sigma_{t}^{2}I),
\end{align}
where $t$ denotes the diffusion steps. The training objective minimizes the error between the predicted and ground-truth noise:
\begin{equation}
\mathcal{L}_{\theta_{av}} = \mathbb{E}_{z_v^t, z_a^t, t, \epsilon_v, \epsilon_a} \left[
\|\epsilon_v - \epsilon_{\theta_{av}}(z_v^t, z_a^t, t)\|_2^2 + \|\epsilon_a - \epsilon_{\theta_{av}}(z_a^t, z_v^t, t)\|_2^2
\right].
\end{equation}

\noindent \textbf{A-V Alignment Module.} To facilitate effective cross-modal interaction, we introduce an \textbf{AV-CrossAttn} module within the DiT blocks. This module shares a similar structure with a multi-head self-attention (MHSA) block but has its own trainable parameters. The audio features $x_a$ and video features $x_v$ are used as queries, keys, and values in a cross-attention mechanism, allowing each modality to attend to the features of the other. This explicit interaction at each denoising step enables the model to learn the intricate correlations between speech and facial movements, resulting in highly synchronized and natural outputs.

{\noindent \bf Entropy Encoding.} We employ a range-based arithmetic coder to compress the quantized tokens based on their predicted probabilities, generating the final bitstream.

\SubSection{Loss and Tuning}
\label{sec:loss}
To ensure high-quality and synchronized output, we employ a multi-component loss function. A lip-sync loss $L_{sync}$ is computed using a pre-trained AV-HuBERT model to measure the cosine similarity between audio and visual (lip region) features, ensuring accurate mouth movements. The overall training objective combines this with a perceptual loss $\mathcal{L}_{per}$, an adversarial loss $\mathcal{L}_{adv}$, and a feature matching loss $\mathcal{L}_{fea}$ to supervise the end-to-end training process:
\begin{equation}
\label{eq_training}
\mathcal{L}_{total} = \lambda_{per} \mathcal{L}_{per} + \lambda_{adv} \mathcal{L}_{adv} + \lambda_{fea} \mathcal{L}_{fea} + \lambda_{sync} L_{sync}
\end{equation}
where the weighting parameters are empirically set to balance reconstruction fidelity, naturalness, and audio-visual synchronization.

\Section{Experimental Results}

\SubSection{Experimental Settings}

\noindent\textbf{Implementation Details.} Our framework is implemented in PyTorch and trained on the VoxCeleb2 dataset~\cite{Nagrani17} using NVIDIA A100 GPUs. For the diffusion process, we utilize pretrained VAEs from Stable Diffusion to encode video frames into a latent space and a pretrained audio VAE for the audio stream. The AVDiff model uses a frozen DiT-XL/2 backbone with lightweight trainable layers for cross-modal attention.

\noindent\textbf{Compared Algorithms.} We compare AVCC against the VVC standard~\cite{bross2021overview} and leading GFVC methods: CTTR~\cite{chen2023compact}, CFTE~\cite{CHEN2022DCC}, FV2V~\cite{FV2V}, and FOMM~\cite{FOMM}. We use 50 test videos from the VoxCeleb testing set, each with 250 frames at $256 \times 256$ resolution.

\noindent\textbf{Evaluation Measures.} We use perceptual metrics DISTS~\cite{dists} and LPIPS~\cite{lpips}, alongside traditional metrics like PSNR and SSIM. Lower DISTS/LPIPS indicate better quality.

\begin{figure*}[t]
\centering
\centerline{\includegraphics[width=1\linewidth]{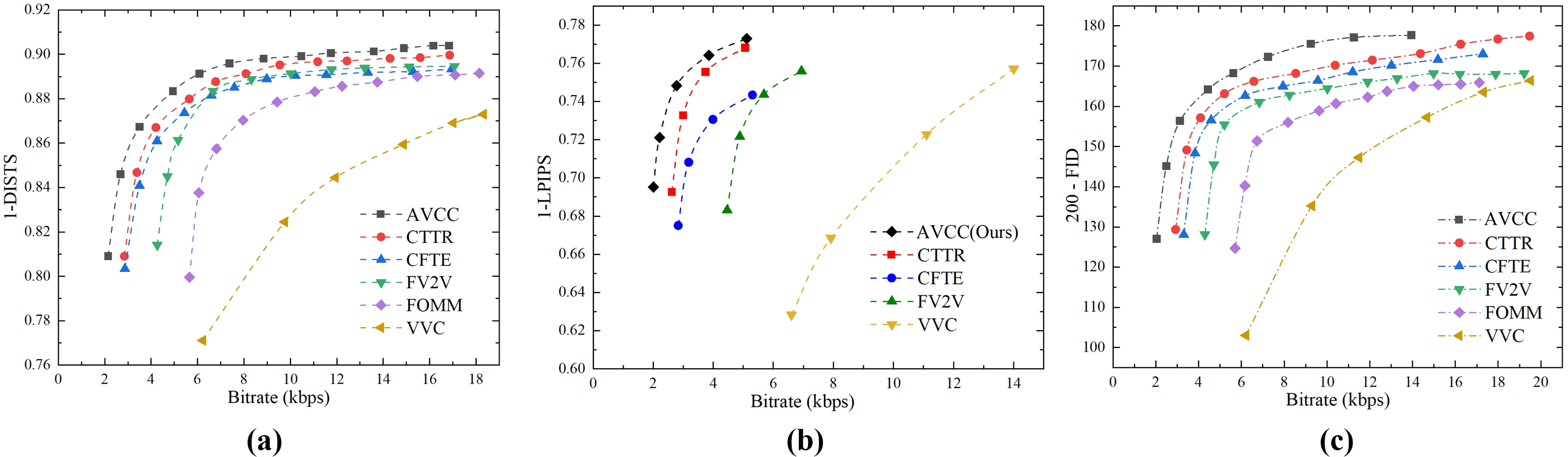}}
\caption{Rate-Distortion (RD) performance comparison. AVCC is benchmarked against VVC and state-of-the-art GFVC methods using three different perceptual metrics: (a) 1-DISTS, (b) 1-LPIPS, and (c) 200-FID. Our method consistently achieves superior reconstruction quality (higher scores) at similar bitrates across all metrics.}
\label{fig:bpp-lpips}
\vspace{-1em}
\end{figure*}

\begin{table}[t]
\scriptsize
\centering
\caption{Video channel BD-Rate savings (\%) of AVCC compared to other GFVC methods. Results are standardized to 256$\times$256 resolution.}
\begin{tabular}{lcccccc}
\toprule
\textbf{Methods} & \multicolumn{2}{c}{\textbf{MEAD}} & \multicolumn{2}{c}{\textbf{Facestar}} & \multicolumn{2}{c}{\textbf{Voxceleb}} \\
\cmidrule(lr){2-3} \cmidrule(lr){4-5} \cmidrule(lr){6-7}
 & \textbf{DISTS} & \textbf{PSNR} & \textbf{DISTS} & \textbf{PSNR} & \textbf{DISTS} & \textbf{PSNR} \\
\midrule
CTTR & 16.6 & 9.8 & 22.1 & 15.4 & 18.8 & 10.7 \\
IFVC & 28.1 & 17.6 & 33.2 & 15.9 & 31.9 & 18.4 \\
FV2V & 52.9 & 33.6 & 58.3 & 38.5 & 49.7 & 31.7 \\
VVC & 93.4 & 5.9 & 96.4 & 9.3 & 94.3 & 7.2 \\
\bottomrule
\label{table:bd_rate_comparison}
\end{tabular}
\end{table}

\begin{table}[h]
\centering
\caption{Audio-channel performance comparison against state-of-the-art audio codecs. MEL/STFT/WER are error metrics ($\downarrow$), while VIS is a quality score ($\uparrow$).}
\scriptsize
\begin{tabular}{lccccc}
\toprule
Model & Kbps & MEL$\downarrow$ & STFT$\downarrow$ & VIS$\uparrow$ & WER$\downarrow$ \\
\midrule
Ground Truth & - & 0.0 & 0.0 & 4.99 & 2.09 \\
\midrule
Encodec \cite{defossez2022high} & 1.50 & 5.39 & 4.47 & 3.04 & 5.0 \\
HiFi-Codec \cite{hifi++} & 2.00 & 4.35 & 3.61 & 3.11 & 3.6 \\
Semantic \cite{liu2024semanticodec} & 1.40 & 4.45 & 3.79 & 3.33 & 3.4 \\
\midrule
AVCC(high) & 1.40 & \textbf{4.43} & \textbf{3.71} & \textbf{3.38} & \textbf{3.22} \\
AVCC(low) & 0.36 & 5.11 & 3.92 & 2.87 & 19.3 \\
\bottomrule
\end{tabular}
\label{tab:audio}
\end{table}

\SubSection{Performance Comparisons}

\noindent\textbf{Rate-Distortion Performance.} Figure~\ref{fig:bpp-lpips} shows the RD performance for video reconstruction. AVCC consistently outperforms all baseline methods, including VVC and other GFVCs, across various perceptual metrics. At very low bitrates (e.g., 2 kbps), our method maintains high perceptual quality (low FID score), whereas competing methods suffer from noticeable artifacts. This demonstrates the effectiveness of our joint audio-visual approach in achieving a better balance between compression rate and visual fidelity. Table~\ref{table:bd_rate_comparison} further confirms these gains, showing significant BD-Rate savings over baselines on multiple datasets.

\noindent\textbf{Audio Performance.} As shown in Table~\ref{tab:audio}, our method also achieves state-of-the-art audio compression performance. At 1.40 kbps, AVCC outperforms specialized audio codecs like Encodec and Semantic Codec on metrics evaluating spectral distance (MEL, STFT), quality (VIS), and word error rate (WER). This is because our model leverages visual cues (lip movements) to reconstruct the audio more accurately, a benefit unavailable to audio-only codecs.

\begin{figure}
    \centering
    \includegraphics[width=0.98\linewidth]{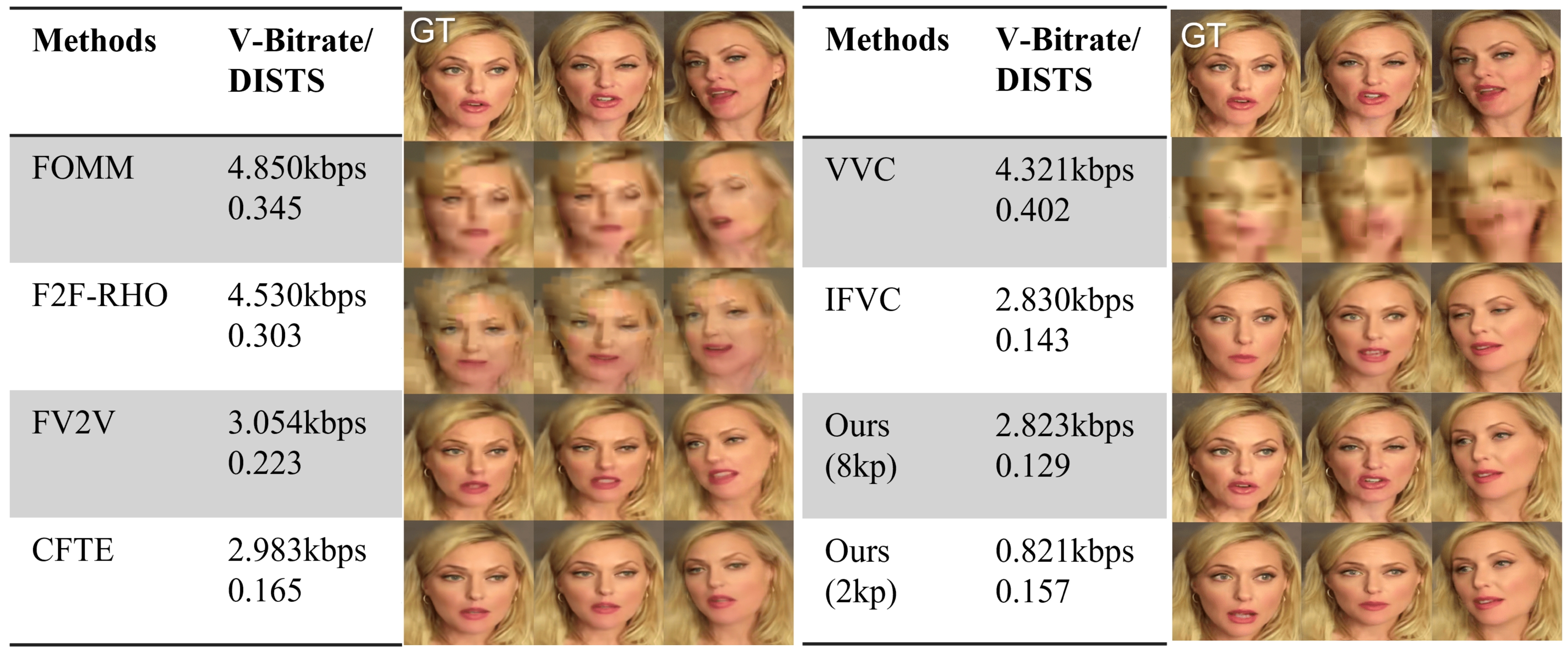}
   \caption{Visual quality comparisons. At an ultra-low bitrate of 2 kbps, our AVCC method preserves facial details and expression more effectively than baseline methods, resulting in better perceptual quality (lower DISTS score).}
\label{fig:sub-rate-dists}
\end{figure}

\begin{figure}
    \centering
    \includegraphics[width=0.9\linewidth]{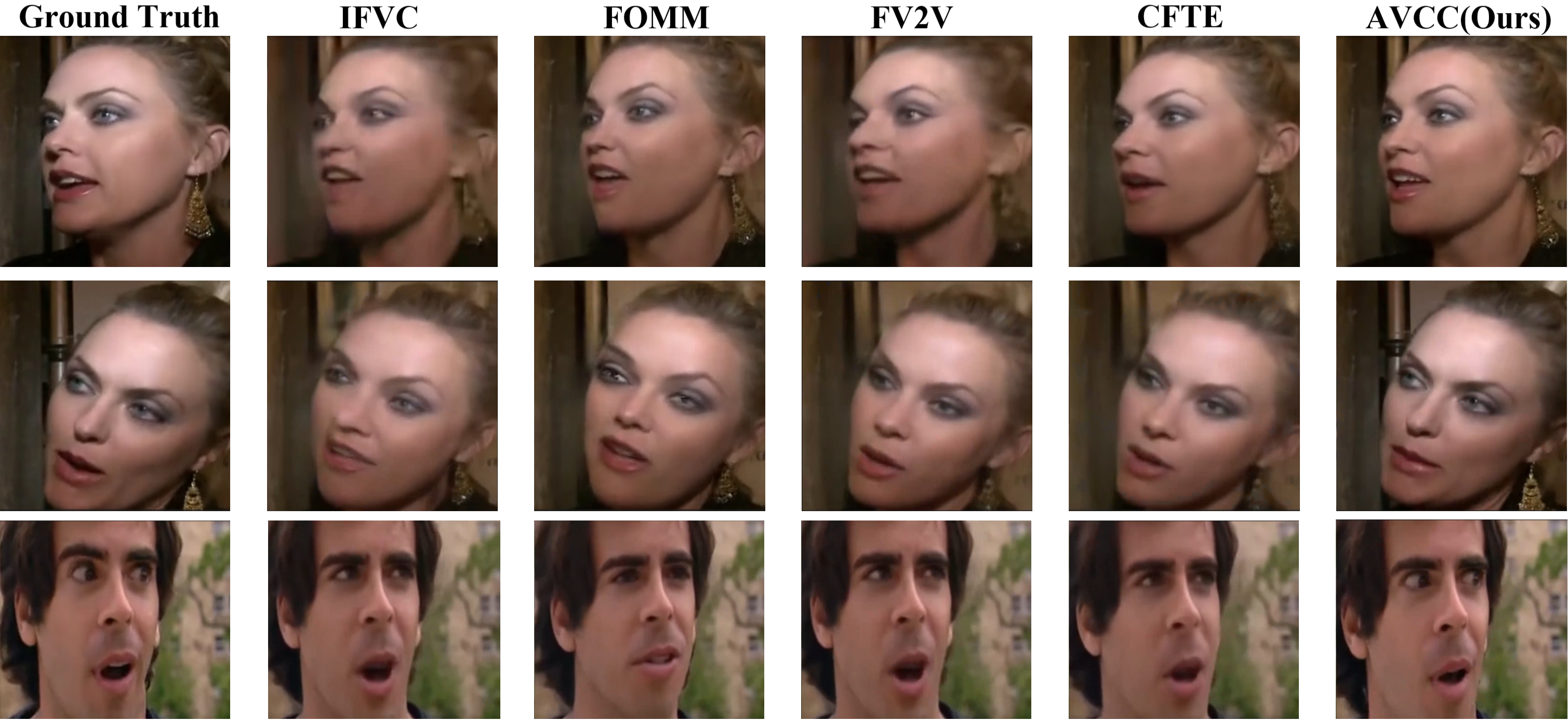}
   \caption{Visual quality comparisons under large-angle rotations. Our method demonstrates superior consistency and visual quality, effectively preserving facial structure and motion dynamics where others might fail.}
\label{fig:extre}
 \end{figure}

\noindent\textbf{Subjective Quality and Extreme Cases.} Visual comparisons in Figure~\ref{fig:sub-rate-dists} confirm the objective results. At ultra-low bitrates, AVCC reconstructs vivid faces with precise expressions, while baselines show clear artifacts. Furthermore, as shown in Fig.~\ref{fig:extre}, our method exhibits exceptional robustness in scenarios with large-angle head rotations. Unlike traditional methods that often fail to maintain consistent facial structure under such conditions, our approach ensures high visual fidelity and temporal coherence, highlighting its ability to handle challenging poses.

\begin{figure}[]
    \centering
    \includegraphics[width=0.99\linewidth]{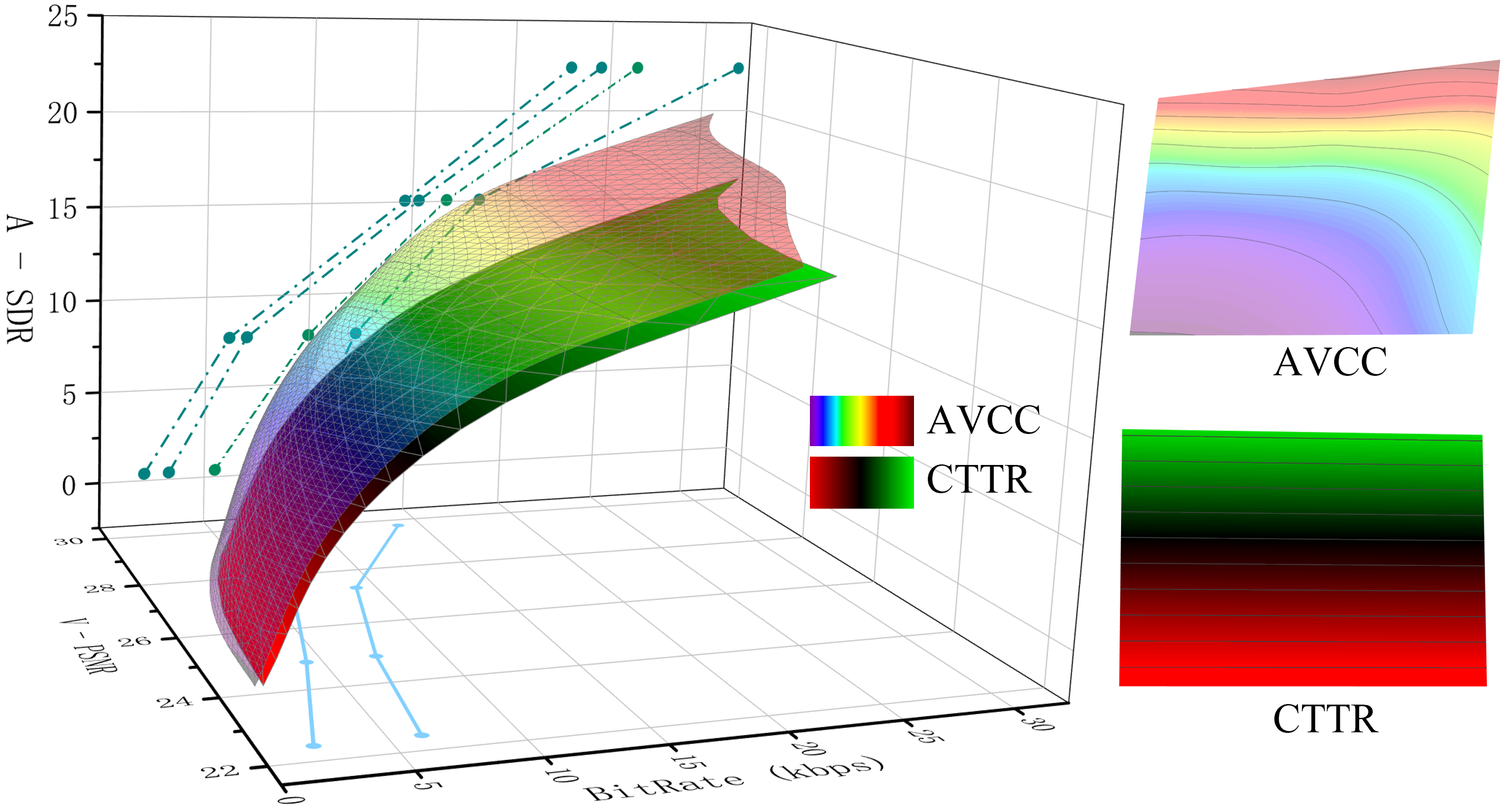}
   \caption{3D Rate-Distortion (RD) surface relating bitrate (X), Video-PSNR (Y), and Audio-SDR (Z). The AVCC surface's contortion indicates a strong correlation between audio and video quality. The contour plots (right) confirm this: CTTR shows no correlation, while AVCC shows a clear positive relationship, especially at low bitrates (red region).}
    \label{fig:RDmesh}
    \vspace{-0.6em}
\end{figure}

\noindent\textbf{Joint Rate-Distortion Analysis.} Figure~\ref{fig:RDmesh} illustrates the 3D Rate-Distortion surface, showing the interplay between the total bitrate, video quality (PSNR), and audio quality (SDR). For CTTR, which compresses modalities independently, the surface is smooth, showing no inter-modal correlation. In contrast, the AVCC surface shows pronounced curvature, demonstrating that the compression quality of both modalities benefits from cross-modal information. The contour plot on the right further highlights this: CTTR shows no correlation, whereas AVCC exhibits a clear positive correlation between audio and video quality, especially at low bitrates. This visually confirms the effectiveness of our joint compression scheme.

\SubSection{Ablation and Complexity Analysis}

\noindent\textbf{Ablation Study.} We conducted ablation studies to validate the key components of our framework. Table~\ref{tb:ablation-a2v} investigates the impact of the AV-CrossAttn module. We compare our full model against a variant where audio-visual conditioning is disabled. The results show that enabling AV-CrossAttn significantly improves performance, especially at lower bitrates. For example, at a video rate of 578 kbps, cross-modal attention improves the FID score by 0.71 and substantially enhances lip synchronization (LipSync score from 3.453 to 6.890). This confirms that leveraging audio information is crucial for reconstructing high-fidelity, synchronized facial movements.

\begin{table}[t]
\centering
\scriptsize
\caption{Ablation study on the AV-CrossAttn module. Comparing performance with ($\checkmark$) and without ($\times$) cross-modal attention at different video bitrates (controlled by token rate).}
\label{tb:ablation-a2v}
\begin{tabular}
{@{}p{1.2cm}p{1.2cm}p{0.8cm}p{0.8cm}p{0.8cm}p{0.8cm}@{}}
\toprule
 \textbf{Tokens/s} & \textbf{AV-CrossAttn} & \textbf{Rate-V}  & \textbf{FID}$\downarrow$ & \textbf{LipSync}$\uparrow$ & \textbf{PSNR}$\uparrow$ \\ 
\midrule
\textcolor{blue}{8} & $\times$ & 1913 & 22.49 & 6.433 & 24.13 \\ 
\textcolor{blue}{4} & $\times$ & 1109 & 22.87 & 4.230 & 23.13 \\ 
\textcolor{blue}{2} & $\times$ & 578  & 23.14 & 3.453 & 20.23 \\ 
\midrule
\textcolor{red}{8} & \checkmark & 1913 & \textbf{22.05} & \textbf{7.279} & \textbf{29.02} \\ 
\textcolor{red}{4} & \checkmark & 1109 & \textbf{22.30} & \textbf{7.123} & \textbf{27.94} \\ 
\textcolor{red}{2} & \checkmark & 578 & \textbf{22.43} & \textbf{6.890} & \textbf{22.41} \\ 
\bottomrule
\end{tabular}
\end{table}

\noindent\textbf{Computational Complexity.} A limitation of our approach is its computational complexity compared to traditional codecs. As shown in Table~\ref{tab:comparison_methods}, the diffusion-based decoder (AVCC-Diffusion) is slower. However, our multimodal encoding framework is versatile. We demonstrate this by substituting the decoder with a traditional 3DMM model (AVCC-3DMM). This configuration achieves a real-time encoding/decoding speed comparable to previous GFVC methods like CTTR, while still maintaining superior quality and bitrate performance due to the efficiency of our joint audio-visual motion representation.

\begin{table}[h]
    \centering
    \caption{Comparison of Methods in Terms of Time Cost (FPS), Perceptual Quality (1-LPIPS), and Bitrate.}
    \begin{tabular}{lccc}
        \toprule
        \textbf{Methods} & \textbf{FPS} & \textbf{1-LPIPS}$\uparrow$ & \textbf{Bitrate (kbps)} \\
        \midrule
        VVC            & 0.67         & 0.619          & 7230 \\
        CTTR           & 0.22         & 0.695          & 3293 \\
        AVCC-Diffusion & 0.13         & \textbf{0.713} & \textbf{2530} \\
        \textcolor{blue}{AVCC-3DMM}      & \textbf{0.23}         & 0.702          & 2854 \\
        \bottomrule
    \end{tabular}
    \label{tab:comparison_methods}
\end{table}

\SubSection{Potential Application Scenarios}
The proposed multi-modal encoding framework supports several advanced applications. In video conferencing, it can ensure continuous transmission by using available modalities to compensate for packet loss—if audio is noisy, video can help predict audio tokens, and vice versa. For live streaming, it can gracefully handle network fluctuations by estimating missing video data from the audio stream. Furthermore, it enables new possibilities in accessibility, such as generating real-time subtitles by leveraging both audio and visual lip-reading cues, and creating realistic virtual avatars whose facial expressions are driven synchronously by either audio or text input.

\Section{Conclusion}
We have introduced AVCC, a novel framework for generative face video coding that, for the first time, integrates audio and video streams into a unified compression pipeline. By leveraging the inherent correlation between audio and visual modalities through a shared diffusion model, our method significantly improves compression efficiency and reconstruction quality, especially in ultra-low bitrate scenarios. Experimental results demonstrate that AVCC surpasses the latest VVC standard and other state-of-the-art GFVC algorithms in both objective and subjective evaluations. This work highlights the potential of cross-modal learning for the future of video communication and provides a strong foundation for the development of more sophisticated multimodal compression systems.

\Section{References}
\bibliographystyle{IEEEtran}
\bibliography{refs}

\end{document}